\documentstyle[aps,prl,amsmath,mathrsfs]{revtex} 
 
\begin{document} 
\draft 
 
\title{Detection of QED vacuum nonlinearities in 
  Maxwell's equations by the use of waveguides} 
 
\author{Gert Brodin$^1$, Mattias Marklund$^2$ and Lennart Stenflo$^1$} 
 
\address{$1$ Department of Plasma Physics, Ume{\aa} University, 
SE--901 87 Ume{\aa}, Sweden} 
\address{$2$ Department of Electromagnetics, Chalmers University of 
Technology, SE--412 96 G\"oteborg, Sweden} 
 
\date{\today} 
 
\maketitle 
 
\begin{abstract} 
We present a novel method for detecting nonlinearities, 
due to quantum electrodynamics through photon--photon 
scattering, in Maxwell's equation. The photon--photon scattering 
gives rise to self-interaction terms, which
are similar to the nonlinearities due to the 
polarisation in nonlinear optics.   
These self-interaction terms vanish in the limit of parallel 
propagating waves,   
but if instead of parallel propagating waves the modes generated 
in waveguides are  
used, there will be a non-zero total effect. Based on this idea, we 
calculate the nonlinear excitation of new modes and estimate the 
strength of this effect. Furthermore, we suggest a principal 
experimental setup.      
\end{abstract} 
\pacs{PACS number(s): 12.20.Ds, 42.50.Vk} 
 
 
According to QED, the non-classical phenomenon of photon--photon scattering 
can take place due to the exchange of virtual electron--positron
pairs. The observation of this scattering is a long-standing
experimental challenge which has yet to be met. Photon-photon scattering
is a second order effect (in terms of the fine structure constant $\alpha 
\equiv e^{2}/2\varepsilon _{0}hc\approx 1/137$), and it can in
standard notation be formulated in 
terms of the Euler--Heisenberg Lagrangian density 
\cite{Heisenberg-Euler,Schwinger} 
\begin{equation}  \label{eq:lagrangian}
  {\mathscr{L}} = \varepsilon_0{\mathscr{F}} + \xi(4{\mathscr{F}}^2 +  
    7{\mathscr{G}}^2) \ , 
\end{equation} 
where  
\[ 
  \xi \equiv \frac{20\alpha^2\varepsilon_0^2\hbar^3}{45m_e^4c^5} \ , 
\]
${\mathscr{F}} \equiv \frac{1}{2}(E^2 - c^2B^2)$, ${\mathscr{G}} 
\equiv c\mathbf{E}\cdot{\bf B}$, and ${\mathscr{F}}^2$ and
${\mathscr{G}}^2$ are the QED corrections. Here $e$ is the electron
charge, $c$ the velocity of light, $h$ the Planck constant and $m_e$
the electron mass. 
We note that ${\mathscr{F}} = {\mathscr{G}} = 0$ in the 
limit of parallel propagating waves. 
The latter terms in (\ref{eq:lagrangian})
represent the effects of vacuum 
polarisation and magnetisation, and the QED corrected Maxwell's vacuum 
equations take the classical form using  
\[
  {\bf D} = \varepsilon_0{\bf E} + {\bf P} \ , \quad  
  {\bf H} = \frac{1}{\mu_0}{\bf B} - {\bf M} \ ,  
\]
where ${\bf P}$ and ${\bf M}$ are of third order in the field amplitudes  
${\bf E}$ and ${\bf B}$, and $\mu_0 = 1/c^2\varepsilon_0$.
Furthermore, they contain terms 
${\mathscr{F}}$ and ${\mathscr{G}}$ such that
${\bf P} = {\bf M} = 0$ in the limit of parallel propagating waves.  
It is therefore necessary to use other waves in 
order to obtain an effect from these QED corrections. 
Several attempts have been 
presented in the literature over the years  
\cite{valluri,Rozanov1998,Rozanov1993,Alexandrov,segev,Ding}, 
where Refs.\ \cite{valluri,Rozanov1998,Rozanov1993,Alexandrov} 
mainly focused
on principal issues, whereas the experimental possibilities for 
detection have been discussed in \cite{segev,Ding}.
Soljacic \& Segev concluded that, using their mechanism \cite{segev}, 
the detection of the QED nonlinearities will be technologically viable 
within ten to fifteen years, provided that the laser power increases 
steadily. In this work we suggest the use 
of waveguides as an experimental setup, something which, as far as the 
authors know, has not been discussed previously for this
purpose. The idea of using a waveguide is to achieve a resonant 
coupling between the parallel propagating waves of different 
frequencies.
We calculate the generated electromagnetic field for 
a rectangular waveguide, using the 
$\mathrm{TE}_{01}$- and $\mathrm{TE}_{10}$-modes as pump waves.
In the proposal for the experimental setup, the waveguide is replaced 
by a cavity, in order to prevent the convective loss of energy, and
thereby maximise the saturated amplitude of the excited mode.
The saturation level due to a finite conductivity of the cavity walls is 
estimated. We find that the excited mode can be detected even for 
moderate levels of the pump mode amplitudes, i.e.\ for field 
strengths that can be supported by the cavity walls.
 
In a medium with polarisation ${\bf P}$ and magnetisation ${\bf M}$ 
the general wave equations for ${\bf E}$ and ${\bf B}$ are  
  \begin{equation}
    \frac{1}{c^2}\frac{\partial^2{\bf E}}{\partial t^2}  
      - \nabla^2{\bf E} =  
      -\mu_0\left[ \frac{\partial^2{\bf P}}{\partial t^2}   
      + c^2\nabla(\nabla\cdot{\bf P})  
      + \frac{\partial}{\partial t}(\nabla\times{\bf M)} \right]    
    \ , \label{WaveE}
  \end{equation}
and 
  \begin{equation}
    \frac{1}{c^2}\frac{\partial^2{\bf B}}{\partial t^2}  
      - \nabla^2{\bf B} = 
      \mu_0\left[ \nabla\times(\nabla\times{\bf M})  
      + \frac{\partial}{\partial t}(\nabla\times{\bf P)} \right]    
      \ . \label{WaveB} 
  \end{equation}
Furthermore, the effective polarisation and magnetisation in vacuum due to 
photon-photon scattering induced by the exchange of virtual 
electron-positron pairs are given by (see, e.g., Ref.\ \cite{segev})
\[
    {\bf P} = 2\xi\left[ 2(E^2 - c^2B^2){\bf E}  
      + 7c^2({\bf E\cdot B}){\bf B} \right]  \ ,
\]
and
\[
    {\bf M} = -2 c^2\xi\left[ 2(E^2 - c^2B^2){\bf B}  
      + 7({\bf E\cdot B}){\bf E} \right] \ . 
\]
Next we consider propagation in a rectangular waveguide with dimensions $x_0$ 
and $y_0$ (i.e. the region $0\leq x\leq x_0,$ $0\leq y\leq y_0$ is vacuum 
surrounded by walls that, as a starting point, are assumed to be perfectly 
conducting.). We assume that the ${\rm TE}_{10}$- and  
${\rm TE}_{01}$-modes act as pump waves with distinct frequencies. To   
lowest order (i.e.\ neglecting the vacuum nonlinearities) the fields 
are   
\begin{mathletters}\label{eq:pump10} 
  \begin{eqnarray} 
    B_{1z} &=& \widetilde{B}_{1z}\,\cos\left( \frac{\pi x}{x_0} \right) 
      \exp[{\rm i}(k_1z - \omega_1t)] + {\rm c.c.} \ , \\ 
    B_{1x} &=& -\left( \frac{{\rm i}k_1x_0}{\pi} 
      \right)\widetilde{B}_{1z}\, \sin\left( \frac{\pi x}{x_0} \right)  
      \exp[{\rm i}(k_1z - \omega_1t)] + \mathrm{c.c.} \ , \\ 
    E_{1y} &=& \left(\frac{{\rm i}x_0\omega_1}{\pi}\right)
      \widetilde{B}_{1z}\, 
      \sin\left( \frac{\pi x}{x_0} \right)  
      \exp[{\rm i}(k_1z - \omega_1t)] + \mathrm{c.c.} \ ,  
\end{eqnarray} 
\end{mathletters} 
together with $\omega_1^2 = k_1^2c^2 + \pi^2c^2/x_0^2$ 
for the ${\rm TE}_{10}$-mode, and  
\begin{mathletters}\label{eq:pump01} 
\begin{eqnarray} 
  B_{2z} &=& \widetilde{B}_{2z}\,\cos \left( \frac{\pi y}{y_0}\right)  
    \exp[{\rm i}(k_2z - \omega_2t)] + {\rm c.c.} \ , \\ 
  B_{2y} &=& -\left( \frac{{\rm i}k_{2}y_0}{\pi }  
    \right)\widetilde{B}_{2z}\,\sin\left( \frac{\pi y}{y_0}\right)  
    \exp[{\rm i}(k_2z - \omega_2t)] + {\rm c.c.} \ , \\ 
  E_{2x} &=& -\left(\frac{{\rm i}y_0\omega_2}{\pi}\right)
    \widetilde{B}_{2z}\,\sin%
    \left( \frac{\pi y}{y_0}\right) \exp[{\rm i}(k_2z - \omega_2t)]  
     + {\rm c.c.} \ ,  
\end{eqnarray} 
\end{mathletters} 
together with $\omega_2^2 = k_2^2c^2 + \pi^2c^2/y_0^2$ 
for the ${\rm TE}_{01}$-mode, where ${\rm c.c.}$ stands for complex 
conjugate. Here we have denoted the wave amplitudes, which to lowest
order are constants, by $\widetilde{B}_{z1}$ and $\widetilde{B}_{z2}$,
respectively.
Substituting the linear expression for the fields into the cubic 
nonlinear terms, we note that there will be perturbations with frequency and 
wavenumber $(\omega_3, k_3)$, where the possible combinations are  
$ 
  (\omega_3, k_3) =  
    (\omega_1, k_1) \ , \   
    (\omega_2, k_2) \ , \  
    (2\omega_1 \pm \omega_2, 2k_1 \pm k_2) \ , \ \text{and} \  
    (2\omega_2 \pm \omega_1, 2k_2 \pm k_1) 
$. 
If there is a small perturbation (e.g.\ of the order of $10^{-15}$) of 
the amplitude of any of the original ${\rm TE}$-modes, it would probably 
be a too difficult task to measure such an effect, whereas the appearance of a 
distinctly new frequency, although with small amplitude, will be easier to 
detect. We therefore concentrate on the two latter combinations, which  
are physically equivalent. Furthermore, if any of these 
combinations satisfies the dispersion relation for a natural mode of the 
waveguide, the amplitude of this resonantly driven mode will be much larger 
than the others. We therefore, for definiteness, choose to consider 
the matching condition   
\begin{equation} 
  (\omega_3, k_3) = (2\omega_1 - \omega_2, 2k_1 - k_2) 
  \ , \label{res-cond}  
\end{equation} 
where the dimensions of the waveguide are assumed to be chosen such as to 
make $(\omega _{3},k_{3})$ a natural mode of the waveguide.  

Using the pump modes (\ref{eq:pump10}) and (\ref{eq:pump01}), 
we can get source terms in the wave equations either for a $\mathrm{TE}_{01}$%
-mode or for a $\mathrm{TM}_{01}$-mode. For the latter case, however, the 
frequency matching conditions and dispersions relations cannot be fulfilled 
simultaneously for real values of all wavenumbers, and thus we consider the 
excitation of a $\mathrm{TE}_{01}$-mode, i.e.\ we let  
\[ 
  \omega _3^2 = k_{3}^{2}c^{2} + \frac{\pi^{2}c^{2}}{y_0^{2}} \ 
  . 
\] 
Thus, using Eqs.\ (\ref{eq:pump10}) and (\ref{eq:pump01}), we can evaluate 
the source terms in Eqs.\ (\ref{WaveE}) and (\ref{WaveB}). 
For a waveguide of finite length fulfilling the boundary condition of no  
incoming wave with frequency $\omega _{3}$ at the waveguide starting 
point $z=0$, we then have spatial growth. The ansatz $B_{3z} = 
\widetilde{B}_{3z}(z)\exp[{\rm i}(k_3z - \omega_3t)] + {\rm c.c.}$  
in (\ref{WaveB}) thus gives a linear spatial growth  
\begin{equation} 
  \widetilde{B}_{3z}(z) =  
    \frac{{\rm i}zV}{2k_3}\widetilde{B}_{1z}^2%
    \widetilde{B}_{2z}^{\ast} \ ,  \label{growth1} 
\end{equation} 
provided $k_3$ is not too small. Here the coupling constant $V$ is
\[
  V \equiv 
  \frac{4\xi}{\varepsilon_0}\left(  
  4\omega_3^2 + \frac{k_3^2c^2}{2} + 2\omega_3\omega_2  
  - \frac{7x_0^2}{2y_0^2}k_1^2c^2 - k_2k_3c^2\right) \ , 
\] 
and the star in (\ref{growth1}) denotes complex conjugation. 
 
When designing parameters for an experiment, 
it might be tempting to choose parameters such that the dispersion relation is 
fulfilled for $k_3 = 0$ in which case we instead get a quadratic 
spatial growth of the amplitude.  
However, the case $k_3 = 0$ is not the most 
interesting choice for two reasons. Firstly, the group velocity of and 
thereby the energy flux of the excited mode is proportional to $k_{3}$. 
Secondly it is impossible for $\omega_3 > \omega_1, \omega_2$ to hold 
when $k_3 = 0$. The reason for requiring the excited wave to have a higher 
frequency than the others is that we then have the possibility to use 
waveguide filtering (see below) in order to measure the excited wave without 
any disturbing signals from the pump waves. Since $k_3$ cannot be small as 
compared to $1/x_0$ for $\omega_3 > \omega_1, \omega_2$ to hold, the 
spatial growth in (\ref{growth1}) may be too slow for practical purposes, 
and we are therefore motivated to consider a cavity rather than a waveguide. 
Since the waves propagating in positive and negative directions are 
identical in that cavity, the boundary conditions imply temporal rather than 
spatial growth. If we assume that all waves have $k > 0$ (are
propagating in the positive $z$-direction) in the waveguide example,
the coupling coefficient in a cavity 
can be found from the waveguide result, simply by noting that the positive 
propagating part of the standing pump waves couples to the positive 
propagating part of the excited standing wave, and vice versa. Since  
$V$ is a quadratic function of the wavenumbers, the same 
coupling strengths apply for standing waves  
$\sim \sin(n\pi z/L)\exp[-{\rm i}\omega t]$,  
where $L$ is the length of the cavity, as for 
propagating waves $\sim \exp [{\rm i}(kz-\omega t)]$, if we just let the 
wavenumber be $k=\pm n\pi /L$ where the sign corresponds to the direction of 
propagation, and $n$ is a positive integer.
For design purposes we must keep in mind, however, that we now 
have additional constraints relating the frequencies and dimensions since the 
wavenumbers cannot be chosen continously. If we let $\widetilde{B}_{z}$ 
represent the standing wave amplitude, and modify the ansatz such that the 
excited ampitude depends on time, we find that the temporal growth in a 
cavity is  
\[
  \widetilde{B}_{3z}(t) = \frac{{\rm i}tc^2V}{2\omega_3}%
    \widetilde{B}_{1z}^2\widetilde{B}_{2z}^{\ast} \ . \label{growth2} 
\]
Saturation occurs when the amplitude is large enough for linear damping due 
to a finite conductivity to balance the driving term. The saturated 
equilibrium amplitude is found to be  
\begin{equation} 
  |\widetilde{B}_{3z{\rm eq}}| =  
    \frac{c^2V}{2\omega_3\Gamma}%
    |\widetilde{B}_{1z}^2||\widetilde{B}_{2z}| \ ,  
  \label{sat-amp} 
\end{equation} 
where $\Gamma$ is the linear damping rate of mode 3. If the linear
damping is due to a finite conductivity $\sigma$ of the walls, we have
\[ 
  \Gamma = U\frac{\omega_3R}{\mu_0c}  \label{damping} 
\]
in the regime $\Gamma/\omega_3\ll 1$,where $\delta \equiv  
{\rm Re}\left[ \left( {\rm i}\mu_0\sigma\omega_3 \right)^{1/2}\right]$  
is the skindepth, $R={\rm Re}(\sigma^{-1})/\delta $ is the surface  
resistance of the cavity walls, and $U$ is a function of the geometry that 
is of the order of unity. 
 
In order to avoid problems due to the large amplitude pump signals when 
trying to measure the excited mode, it is convenient to use a slightly 
modified cavity rather than an idealized cubic cavity. Such a cavity
could consist of two parts; one cavity (I), in which the signal is
generated, attached to another cavity (II), acting as a waveguide
filter. The 
dimensions of cavity I should be chosen such as to keep the frequency 
of the excited mode above cut-off in cavity II, whereas the pump modes should 
be below cut-off. By measuring the excited signal far enough into
cavity II 
(in practice 30-40 pump decay lengths), the pump waves effectively vanish, 
and we can forget about the disturbing influence of the pump waves. However, 
note that a certain fine tuning of the length of cavity II might be necessary 
to keep a maximally efficient phase relation between the 
excited eigenmode and the pump waves in cavity I. 
 
Next we demonstrate that the experimental model setup gives signals 
that can be detected with presently available technology. High performance, 
i.e.\ large electromagnetic fields combined with low dissipative losses, can 
be found in superconducting cavities, which among other things are 
used for particle accelerator purposes \cite{cavity performance}. 
Adopting data from these 
experiments, we assume that the pump waves have a field-strength  
$E_{\rm crit}\sim 30\,\mathrm{MV/m}$, i.e. close to the maximum that 
can be tolerated by the walls without field emissions. For a cavity 
resistance $R \sim 1\,{\rm n}\Omega$, corresponding to  
superconducting niobium at a temperature $1.4 \, \mathrm{K}$ and a  
frequency $\omega_3 \sim 2\times 10^{10}\,\mathrm{rad/s}$,  
we find from Eq. (\ref{sat-amp}) that the saturated energy  
flux $P_3$ of mode 3 is of the order of $P_3 \sim 
10^{-6}\,\mathrm{W/m}^{2}$.   
Here we have used the simple order of magnitude estimate that all 
wavelengths are comparable to the dimensions $x_0$ and $y_0$ when 
evaluating $V$. Clearly this energy flux is above the  
detection level by 
several orders of magnitude. Note, however, the importance of the 
superconducting walls for the output level of the excited mode. For copper 
at room temperature, the cavity resistance increases by a factor 
 $\sim 10^7$ as compared to the above example, and consequently the 
energy flux of the excited mode falls by a factor $\sim 10^{-14}$. In 
this latter case it is questionable whether the excited signal can be 
detected.  
 
To our knowledge, it has not been possible to verify the Euler--Heisenberg 
Lagrangian density experimentally. The above calculation of the QED mode 
coupling strength and the subsequent estimations suggest that it can be very 
suitable to use two pump modes in a superconducting cavity for this purpose. 
The parameters of the problem should be designed such as to simultaneously 
fulfill the dispersion relations of each mode together with the matching 
condition (\ref{res-cond}). Naturally, care must be taken when 
drawing the conclusions, since there are certain effects that we have not 
yet addressed. For example, in the present model the conductivity of the 
walls is linear, but in principle there might be a nonlinear contribution to 
the conductivity that could give rise to a small signal at the same 
frequency as the QED contribution. On the other hand, to our knowledge, 
there are no theoretical or experimental reports of such effects. 
Secondly, in reality the vacuum in the waveguide will not be perfect, and in 
principle this may lead to dielectric breakdown for the pump field strength 
considered in the above estimate. However, we do not expect this particular 
effect to be a serious threat against our proposal, since similar electric 
field strengths have been reached in present experiments \cite{cavity 
performance}, and the pump field strength that is actually needed is much 
less than the one considered in the estimate. Thus we conclude that it is 
likely that the effect of photon--photon scattering in vacuum due to the 
exchange of virtual electron--positron pairs can be meaured using existing 
technology.  
 
 
 
\end{document}